\documentclass
[namedreferences]{SolarPhysics}
\usepackage[optionalrh]{spr-sola-addons} 
\usepackage{graphicx}                    
\usepackage{color}                       
\usepackage{url}                         

\newcommand{\etal}{{\it et al.}}

\newcommand{\kms}{km~s\ensuremath{^{-1}}}
\newcommand{\arcs}{\mbox{\ensuremath{^{\prime\prime}}}}

\newcommand{\adv}{    {\it Adv. Space Res.}}

\newcommand{\aap}{    {\it Astron. Astrophys.}}
\newcommand{\aaps}{   {\it Astron. Astrophys. Suppl.}}

\newcommand{\ag}{     {\it Ann. Geophys.}}

\newcommand{\apj}{    {\it Astrophys. J.}}
\newcommand{\apjl}{    {\it Astrophys. J. Lett}}

\newcommand{\grl}{    {\it Geophys. Res. Lett.}}

\newcommand{\jgr}{    {\it J. Geophys. Res.}}

\newcommand{\pasj}{   {\it Pub. Astron. Soc. Japan}}

\newcommand{\solphys}{{\it Solar Phys.}}

\newcommand{\ssr}{    {\it Space Sci. Rev.}}

\begin{document}
\begin{article}
\begin{opening}
\title{Signatures of the slow solar wind streams from active regions in the inner corona}
%

\author{V.~\surname{Slemzin}$^{1}$\sep
        L.~\surname{Harra}$^{2}$\sep
        A.~\surname{Urnov}$^{1}$
        S.~\surname{Kuzin}$^{1}$
        F.~\surname{Goryaev}$^{1, 3}$
        D.~\surname{Berghmans}$^{3}$
       }
%

%
   \institute{$^{1}$ P.N. Lebedev Physical Institute, Leninsky Pr., 53, Moscow, 119991, Russia.
                     email: \url{slem@sci.lebedev.ru}\\
              $^{2}$ UCL-Mullard Space Science Laboratory, Holmbury St Mary, Dorking, Surrey, RH5 6NT, UK.
                     email: \url{lkh@mssl.ucl.ac.uk} \\
                     $^{3}$ Royal Observatory of Belgium, Ringlaan 3, 1180 Brussels, Belgium.
                     email: \url{david.berghmans@oma.be} \\
             }

\begin{abstract}

Identification of the solar wind sources is an important question in solar physics. The existing solar wind  models (e.g. Wang-Sheeley-Arge model) provide the approximate locations of the solar wind sources based on magnetic field extrapolations. It has been suggested recently that plasma outflows observed at the edges of active regions may be a source of the slow solar wind. To explore this we analyze an isolated active region (the AR) adjacent to small coronal hole (CH) in July/August 2009. On August 1 Hinode/EUV Imaging Spectrometer observations showed two compact outflow regions in the corona. Coronal rays were observed above the ARCH region on the eastern limb on July 31 by \textit{STEREO-A} /EUVI  and at the western limb on August 7 by \textit{CORONAS-Photon}/TESIS telescopes. In both cases the coronal rays were co-aligned with open magnetic field lines given by the Potential Field Source Surface model, which expanded into the streamer. The solar wind parameters measured by \textit{STEREO-B}, \textit{ACE}, \textit{WIND}, \textit{STEREO-A} confirmed the identification of the ARCH as a source region of the slow solar wind. The results of the study support the suggestion that coronal rays can represent signatures of outflows from ARs propagating in the inner corona along open field lines into the heliosphere.

\end{abstract}

%
\end{opening}
%
\section{Introduction}

Since the discovery of the solar wind, identification of its sources at the Sun has the principal meaning for fundamental solar physics as well as for practical forecasting of the space weather. It is conventional now that the fast solar wind dominating at solar minimum originates from polar coronal holes (PCH) \cite{Levine77,Gosling95,McComas01}. The source regions of the slow solar wind are not so well established. Analysis of {\it Ulysses} and {\it ACE} data suggests that in different phases of the solar cycle the slow wind emanates from various local sources such as boundaries of PCH and equatorial coronal holes (ECH) \cite{Neugebauer98,Neugebauer02,Liewer03}, non-coronal hole sources \cite {Zhao09}, helmet streamers \cite{Sheeley97} and active regions (ARs) \cite{Hefti00,Liewer01,Liewer04}. These identifications were based on juxtaposition of the solar wind parameters measured near the Earth with different solar structures spatially distributed over the Sun under assumptions that propagation of the plasma flows in the inner corona can be described by the potential current-free magnetic field model and that the solar wind in the heliosphere moves along the Archimedian spiral with constant radial speed.

Currently two models are practically used for identification of the solar wind sources and prediction of its parameters near the Earth. (1) The Wang-Sheeley-Arge (WSA) model \cite{Wang90,Arge00} is based on the concept that the main sources of the solar wind are coronal holes associated with rapidly expanding open magnetic fields calculated using the PFSS (Potential Field Source Surface) approximation \cite{Altschuler69,Schatten69,Wang92, Schrijver03}. The initial magnetic field distribution at the solar surface is taken from synoptic maps of the line-of-sight photospheric magnetic field which are constructed from magnetograms periodically provided by on-ground and space-based magnetographs. Between the solar surface and conditional "source surface" (SS) located at $R = 2.5 R_{\mathrm{sun}}$ the coronal magnetic field is assumed to be fully current-free. Then, between SS and outer surface at $21.5 R_{\mathrm{sun}}$ the field is calculated according the Schatten current sheet model \cite {Schatten71}. In the heliosphere the solar wind is supposed to flow along radially directed magnetic field lines. Its velocity is determined by empirical relation with magnetic field expansion factor taking as additional parameter the distance between footpoints of open field lines and the nearest boundary of coronal hole \cite{Arge03}. A simple scheme is used to account for possible interactions of the slow and fast streams between the Sun and the Earth. The WSA model gives a reliable prediction of the solar wind velocity agreed with observations within 10-15$\%$ \cite{Arge00}, especially for the high speed wind from large size polar and middle-latitude coronal holes.(2) The combined WSA-ENLIL model was developed based on the WSA approach in the region between the solar surface and the current sheet and the time-dependent 3D MHD ENLIL model of the heliosphere \cite {Toth96}.

However, application of these models to localization of solar energetic particle (SEP) bursts \cite{Wang06,Nitta06,MacNeice11} has shown that their  accuracy in association of open field lines with physically presumed compact sources at the Sun is limited to 10 -- 20$^0$ in longitude as well as in latitude. The uncertainty in positioning of the sources is primarily produced by inadequacy of the potential magnetic field approximation in the inner corona \cite{Yeates10}, in particular, in the vicinity of ARs where the current-free assumption may be violated. The second source of uncertainty is tardiness of initial synoptic magnetic maps due to their quasi-static nature and unknown evolution of the magnetic field on the hidden side of the Sun \cite{Nitta08}. By estimation of \inlinecite{Rust08}, even if the input magnetic data are updated frequently, the PFSS model only succeed in $\sim$~50\% of cases to identify the coronal segment of open fields. Due to these difficulties as well as complicated variation of the solar wind parameters during its passing through the heliosphere, the direct identification of the local solar wind sources by only large-scale modeling seems to be problematic, and one needs to use additional physical considerations.

Analysis of ionic composition and freezing-in temperature determined from the ionic state distribution in the solar wind near the Earth, helps to identify the principal type of the solar wind sources. Numerous studies have shown that freezing-in temperature (or corresponding O$^{7+}$/O$^{6+}$ ratio) and the FIP (first ionization potential) bias estimated from the iron to oxyden (Fe/O) ratio have different values for various solar wind sources such as streamers (see e.g. \opencite{Raymond97}; \opencite{ Strachan02}; \opencite{Bemporad03}), coronal holes, ARs  and their interfaces (\opencite{Liewer04}; \opencite{Neugebauer02}; \opencite{Ko06}). \inlinecite{Wang09} concluded that the O$^{7+}$/O$^{6+}$ and  Fe/O ratios have the highest values in the slow wind originating from small holes located in and around ARs, the intermediate values correspond to the sources located at the boundaries of large coronal holes (in particular, the polar coronal holes) and the lowest values correspond to the high-speed streams emerging from interiors of large low-latitude holes and equatorwards extensions of the polar coronal holes.

One of presumed sources of the solar wind are plasma outflows often detected in the vicinity of ARs. \inlinecite{Uchida92} reported detection of outflows from ARs observed in X-rays as continual expansion of plasma from hot loops. \inlinecite{Svestka98} observed fans of transient coronal rays above ARs after flares and suggested that they may contribute to the solar wind. In both cases outflows lasted for periods from several hours to several days. \inlinecite{Winebarger01} detected brightness variations along a bundle of coronal loops in the {\it TRACE} 171~\AA\ images and interpreted this phenomenon as a signature of outflows with velocities between 5 and 20 \kms.

Outflows and downflows were detected in active regions by analysis of the EUV Dopplergrams obtained with the spectrometers CDS (Coronal Diagnostics Spectrometer -- \opencite{Harrison97}) and SUMER (Solar Ultraviolet Measurements of Emitted Radiation -- \opencite{Wilhelm97}) onboard the \textit{SOHO} observatory. \inlinecite{Brekke97} observed with CDS significant flows of plasma in AR loops which were interpreted there as  ``siphon flows'' reaching velocities of $\sim$50 \kms\ and more both in the corona and in the transition region. \inlinecite{Marsch04} studied the Doppler shifts of the EUV transition region lines along closed loops in several ARs measured with SUMER in connection with the force-free extrapolation of photospheric magnetic field.

Since {\it Hinode} \cite{Kosugi07} was launched, there has been an increasing interest in studying strong outflows from the edges of ARs. \inlinecite{Sakao07} analyzed a sequence of soft X-ray images taken by the XRT (X-Ray Telescope) instrument \cite{Golub07} on February 20, 2007 and detected moving brightness disturbances in a fan-like structure at the edge of an AR located adjacent to a coronal hole. They interpreted this phenomenon as a signature of continuous outflow of plasma with a temperature of 1.1~MK propagating along open magnetic field lines with the velocity of $\sim$100 \kms\ and suggested that this outflow was a possible source of the solar wind. Using EIS (EUV Imaging Spectrometer - \opencite{Culhane07}), \inlinecite{Harra08} analyzed the Fe{\sc XII} emission line in the same region and found outflow speeds of around 50 \kms, which after correction for projection effect, reached speeds similar to those seen in the XRT imaging data.

The creation and maintenance of the quasi-stationary outflowing plasma is still being debated. There have been various suggestions as to what could create this outflowing plasma. \inlinecite{Harra08} showed through magnetic field extrapolation that the edges of the active regions can reconnect with bipoles lying further away from the ARs. \inlinecite{Baker09} suggested that outflows may originate from specific locations of the magnetic field topology where field lines display strong gradients of magnetic connectivity, namely, quasi-separatrix layers (QSLs). Another option to explain the outflows was put forward by \inlinecite{Murray10}, who studied the special case of an active region inside a coronal hole. In this case they found through simulations that the outflow could be replicated through the process of compression. Some other mechanisms can produce transient or quasi-periodic flows: jet or spicule-like events \cite{McIntosh09} or coronal mass ejection providing outflow from dimming region \cite{Harra10}. The main issue is how to distinguish outflows that actually leave the Sun from those which are just parts of closed ``coronal circulation'' of the solar plasma \cite{Marsch08}. Obviously, the flows contributing to the solar wind should have specific large-scale signatures in the inner corona bounded with their source regions and co-aligned with open magnetic field lines.

A structure of the inner corona has not been studied extensively so far. As stray light near the limb in white-light (WL) is much brighter than the coronal emission, the existing  WL coronagraphs typically observe the corona with reasonable resolution far above the limb, e.g. at the distance of $2.2 R_{\mathrm{sun}}$ from the solar center in the LASCO C2 coronagraph (\opencite{Brueckner95}), or with quickly degrading imaging quality closer to the Sun at $1.3 R_{\mathrm{sun}}$ in {\it STEREO} COR1 coronagraphs (\opencite{Thompson03}), $1.1 R_{\mathrm{sun}}$ in the Mauna Loa MK4 coronagraph (\opencite{Burkepile05}). Rare high-resolution pictures of the whole corona taken during total solar eclipses (e.g. \opencite{Wang07}) show the corona near the limb as broad diffuse structure accumulating multitude of flows distributed over large areas at the Sun irrespective of their temperature. The contrast enhancing techniques (see e.g. \opencite{Druckmuller09}) reveal small local density gradients but do not help much to identify the origins of the coronal streams. The EUV imaging is better capable to localize coronal structures near the limb as well as at large distances, because brightness of the coronal radiation in EUV typically exceeds or comparable with that of stray light at large distances, meanwhile, the spatial resolution remains the same as at the disk. In EUV the coronal radiation is formed from spectral lines of ions excited by collisions and by resonant scattering. Close to the Sun the collisional excitation dominates, hence, brightness of the coronal features is proportional to the squared density indicating local structures with enhanced density in the regions of strong magnetic field condensations.

The first EUV observations of the extended corona in the 175 \AA\ band, collecting the Fe~{\sc ix}--Fe~{\sc xi} lines, up to the radial distances $R=2-3 R_{\mathrm{sun}}$ were carried out with the SPIRIT (SPectrographIc X-Ray Imaging Telescope-spectroheliograph) telescope aboard the {\it CORONAS-F} (Complex Orbital neaR-Earth ObservatioNs of Activity of the Sun) satellite \cite{Slemzin08} at maximum of the 23$^{d}$ solar cycle. These observations discovered existence of large-scale structures -- coronal rays located above some ARs and followed them as the Sun rotates. In the beginning of the 24$^{th}$ solar cycle, in 2009, coronal rays were observed with the TESIS (TElescopic Spectroheligraphic Imaging System telescope) in the 171 \AA\ band (the Fe~{\sc ix}--Fe~{\sc x} lines, \opencite{Kuzin09}). Currently coronal rays are systematically observed with SWAP (Sun Watcher using Active Pixel System detector and imaging processing telescope) \cite{Berghmans06} onboard the {\it PROBA2} (PRoject for OnBoard Autonomy) satellite in the 174 \AA\ band. It is reasonable to suppose that the large-scale coronal rays above ARs may represent coronal signatures of outflows.

On their periphery, ARs often display fan-like structures which are believed to be partially observed high cool loops. \inlinecite{Schrijver99} classified fan structures as long lived loops with the temperature of 1~MK fanning out of the {\it TRACE} field of view. \inlinecite{Berghmans99} reported a detection of propagating disturbances with velocity of the order of 100 \kms\ along partially open fan loops observed both in EIT 195 \AA\ and {\it TRACE} 171 \AA\ images. However, they interpreted this phenomenon as sonic perturbations. Using {\it SOHO}/CDS, \inlinecite{DelZanna03} have confirmed that the loops seen in the {\it TRACE} images have temperature of 0.7--1.1 MK. \inlinecite{Brooks11} found that fan loops observed with {\it Hinode}/EIS and {\it SDO}/AIA (Atmospheric Imaging Assembly onboard Solar Dynamic Observatory) have peak temperatures in the similar 0.8--1.2 MK range.

On the contrary, \inlinecite{Ugarte09} have analyzed with {\it Hinode}/EIS brightness variations in different loop populations and suggested that the fan structures visible in the {\it TRACE} 171 \AA\ channel belong to the same population of cool peripheral loops which are prominent in the Si {\sc vii} and Mg {\sc vii} 0.6 MK lines and develop downflows with velocities in the range 39--105 \kms. The authors of the former publication did not found any evident relation between these fan loops and hotter blueshifted structures which showed transient propagating disturbances seen in the blue wing emission of the Fe~{\sc xii} 195.12 \AA\ line. \inlinecite {Warren11} found that the bright fan-like structures seen in colder transition region lines are dominated by downflows in contrast to outflows seen in the emission lines of the Fe~{\sc xi}--Fe~{\sc xv} ions. The temperature structure of fan loops and their relation to outflows are still in debate. Recently \inlinecite{Young12} suggested, that the fan loops consist of at least two groups of ``strands'': one is hot and stationary, the other is cooler and downfloading. The second strand may represent a later stage of evolution of one or more hotter strands.

A generic relationship between various coronal structures and plasma outflows may be established from the analysis of their temperature distributions with the use of the Differential Emission Measure (DEM) method. In a number of recent works the DEM analysis was applied to explore different regions of the Sun observed by the EIS instrument. The temperature structure of quiet Sun (QS) regions was analyzed in the papers by \inlinecite{Matsuzaki07}, \inlinecite{Warren09} and \inlinecite{Brooks09}. In typical quiet regions and ARs \inlinecite{Matsuzaki07} found out a group of loops with temperatures of $\sim$1~MK and $\sim$2~MK superposed in the line of sight and the low temperature component at $\sim$0.4~MK. \inlinecite{Warren09} presented an analysis of temperature and density measurements above the limb in the quiet corona using the EIS data. \inlinecite{Brooks09} carried out a DEM analysis of the quiet solar corona on the disk. For both cases the authors found the plasma temperature component peaked near 1~MK. \inlinecite{Landi09} reported on a cold, bright portion of an AR observed by the EIS instrument. The emitting region was characterized by a large maximum at $\log T\approx$~5.6, corresponding to transition region temperatures, and a broad tail in the DEM distribution extending to higher temperatures. \inlinecite{Brooks2011} have studied the outflow regions of AR 10978 and determined, in particular, that the electron density and temperature in the outflows had values of $\log N_e = 8.6$ and $\log T = 6.2$ correspondingly. Summing all cited results, one may conclude that typically plasma in outflow regions, fan loops, quiet regions and quiet closed loops have the dominating temperature components in the range $T \sim$1--2~MK.

The aim of this work is to investigate the link between outflows from ARs and coronal structures seen in the EUV range -- coronal rays and fan loops as probable signatures of the outgoing plasma streams. We describe the properties of coronal rays at solar maximum and at solar minimum. We examine the relationship between outflows, coronal structures and the corresponding solar wind for an isolated AR observed with various space solar instruments during a half of the solar rotational period in July--August 2009 at the disk as well as at both limbs. We analyze the main properties of outflows using the EIS intensity and velocity maps in different spectral lines, study the temperature structure of the emitting plasma in different places based on the DEM analysis, and compare coronal signatures of outflows with  magnetic field configurations computed with the use of the PFSS model. Finally, we compare positions of the outflows with the map of solar wind source regions determined using the WSA model and analyze corresponding solar wind data obtained at 1 AU by PLasma And Supra-Thermal Ion Composition Investigation (PLASTIC) instruments on Solar Terrestrial Relation Observatories ({\it STEREO}) Behind and Ahead, Solar Wind Experiment Proton and Alpha Monitor (SWEPAM) onboard Advanced Composition Explorer ({\it ACE}), Solar Wind Experiment (SWE) onboard {\it WIND} spacecrafts. Summary and conclusions section will resume the results of the study.

\section{Coronal rays in the inner corona at solar maximum and minimum}

Ordinary EUV telescopes can observe the solar corona only to the distances $R<1.3-1.5 R_{\mathrm{sun}}$ due to the limited field of view (FOV) and insufficient sensitivity to exponentially decreasing coronal brightness. SPIRIT was the first EUV telescope to have a  coronagraphic mode, in which the direct disk radiation was suppressed by application of an external occulter. It significantly decreases straylight allowing to study a large part of the inner corona from the solar surface to the radial distance of 2.5--3$R_{\mathrm{sun}}$  \cite{Slemzin08}. The successor of SPIRIT, the TESIS EUV telescope \cite{Kuzin09}, which operated in February--November 2009, observed the inner corona using a tiltable mirror and composition of images with different exposure times from tens of seconds to 600 s. Both SPIRIT and TESIS had very low straylight due to their one-mirror Herschel optical design. The SWAP EUV telescope launched on the \textit{PROBA 2} satellite on November 2, 2009 has a design of one-quarter of the EIT telescope with one spectral channel 174 \AA\ \cite{Defise07}. The telescope is able to observe the whole inner corona in a special ``paving'' mode, when the satellite tilts the main axis on $\sim$10~arcmin sequentially in 4 positions (North-East, South-East, North-West and South-West). In each position the telescope takes 80--90 images with a cadence of about 30 s. The images in each paving position are summed and then combined together in the mosaic image of the solar disk and the inner corona. Typically we can see the signal up to 2$R_{\mathrm{sun}}$ from the disk center. The main limitations are the level of straylight and dark current noise (the latter is seriously reduced in the last 1.3 version of the SWAP fits files). The straylight is extrapolated from radial brightness distribution in periphery regions of the image above $R > 2 R_{\mathrm{sun}}$ and then subtracted.

        \begin{figure}    
   \centerline{\includegraphics[width=0.9\textwidth,clip=]{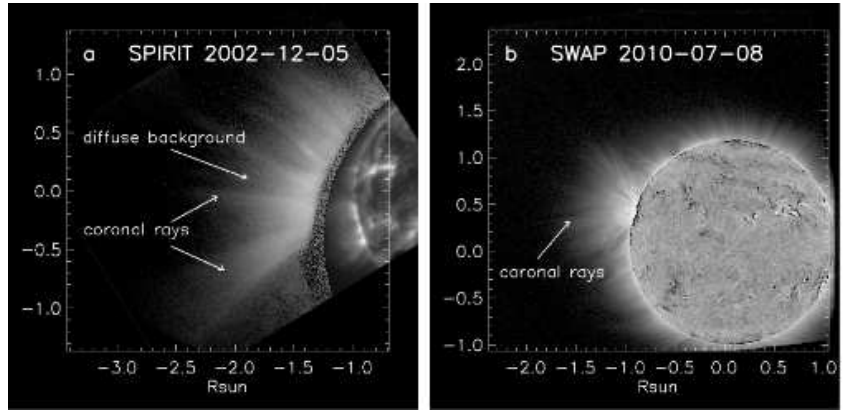}
              }
              \caption{Images of the inner corona in the SPIRIT 175 \AA\ band (Fe~{\sc ix} -- Fe~{\sc xi} lines) taken at solar maximum on December 5, 2002 (a) and in the SWAP 174 \AA\ band (Fe~{\sc ix} -- Fe~{\sc x} lines) taken at starting phase of the 24$^{th}$ solar cycle on July 8, 2010 (b). The distances are counted out from the solar center in units of the solar radius.
                      }
   \label{fig1}
   \end{figure}

Figure \ref{fig1} demonstrates images of the inner corona in the 175 \AA\ band (Fe~{\sc ix} -- Fe~{\sc xi} lines) taken by SPIRIT on December 5, 2002 near the maximum of the 23$^{d}$ solar cycle (a) and by SWAP  in the 174 \AA\ band (Fe~{\sc ix} -- Fe~{\sc x} lines) on July 8, 2010 at very low solar activity at the beginning of the new 24$^{th}$ cycle (b). The SPIRIT image shows that the corona at the eastern limb between the latitudes from $-30^{0}$ to $+60^{0}$ contains a diffuse (non-resolved) background and several bright rays expanding to the distances more than 2.5$R_{\mathrm{sun}}$ from the solar center. Most of the rays are narrow ($\sim$2 -- 3$^{0}$) quasi-radial structures except the ray which expands super-radially in the southeastern direction with the angular spread of about $\sim$30$^{0}$  declined poleward. At the distances between 1.5 and 2$R_{\mathrm{sun}}$ the rays have brightnesses 2--4 times higher than that of the surrounding diffuse corona, below this distance the rays merge with bright loops. Continuous SPIRIT observations during a week in June and December 2002 have shown that coronal rays originated from active regions and followed them as the Sun rotated \cite{Slemzin08}.

The structure of the corona in the SWAP image taken on July 8, 2010 differs from that at solar maximum. It consists of short (less in length than 0.5$R_{\mathrm{sun}}$) fine rays uniformly distributed over the limb and a bundle of longer rays (as long as 1.8--2$R_{\mathrm{sun}}$) expanding from a single AR at the eastern limb. The bundle consists of thin threads super-radially expanding from the AR within the angle $\pm 60^{0}$ with respect to the local solar surface normal.

The nature of coronal rays is still questionable. Phenomenologically, they represent extended plasma structures with enhanced density having the largest length and brightness relative to background in the 1 MK coronal EUV lines of the Fe~{\sc ix} -- Fe~{\sc xi} ions. The luminosity of coronal rays is provided by the mechanisms of electron-ion collisions near the Sun and resonant scattering of the background coronal radiation at larger distances where the electron density exponentially decreased. Some of the rays may be visible parts of large loops disappearing in the middle part due to plasma expansion or variation of its temperature. If the rays contain accelerating plasma, their visible length may be limited by the Doppler dimming effect when the plasma mass velocity achieved some critical value. The real open structures may be identified if they are co-aligned with open magnetic field lines and joined with streamers above SS.

It is important to note that images of the corona obtained during eclipses in the monochromatic Fe~{\sc x} 6374 \AA\ and Fe~{\sc xi} 7892 \AA\ WL lines \cite {Habbal07} displayed similar features as observed by SPIRIT, TESIS and SWAP in the EUV band. These lines are irradiated by the same Fe~{\sc ix} -- Fe~{\sc xi} ions at the plasma temperature of $\sim$1~MK which is responsible for coronal EUV emission. In particular, during the eclipses of 2006, 2008 and 2009 images of the corona  in the 6374 \AA\ and 7892 \AA\ lines contained structures extending up to the distances of 3$R_{\mathrm{sun}}$. Meanwhile, such structures were not observed in the Fe~{\sc xiii} 10747 \AA\ and Fe~{\sc xiv} 5903 \AA\ lines which correspond to the hotter plasma with $T>$~1.2~MK. The authors found that images of the corona in the red 7892 \AA\ line contained localized intensity enhancements at heights ranging from 1.2 to 1.5$R_{\mathrm{sun}}$ with no counterparts in broad band WL images. The conclusion was that the extended coronal structures observed in the Fe~{\sc xi} visible lines originate from solar sources with electronic temperature of 1.1~MK which agrees with our suggestion based on the imaging of the corona in the EUV spectral band.

\section{Case study of the AR in July-August 2009}

\begin{figure}    
   \centerline{\includegraphics[width=0.8\textwidth,clip=]{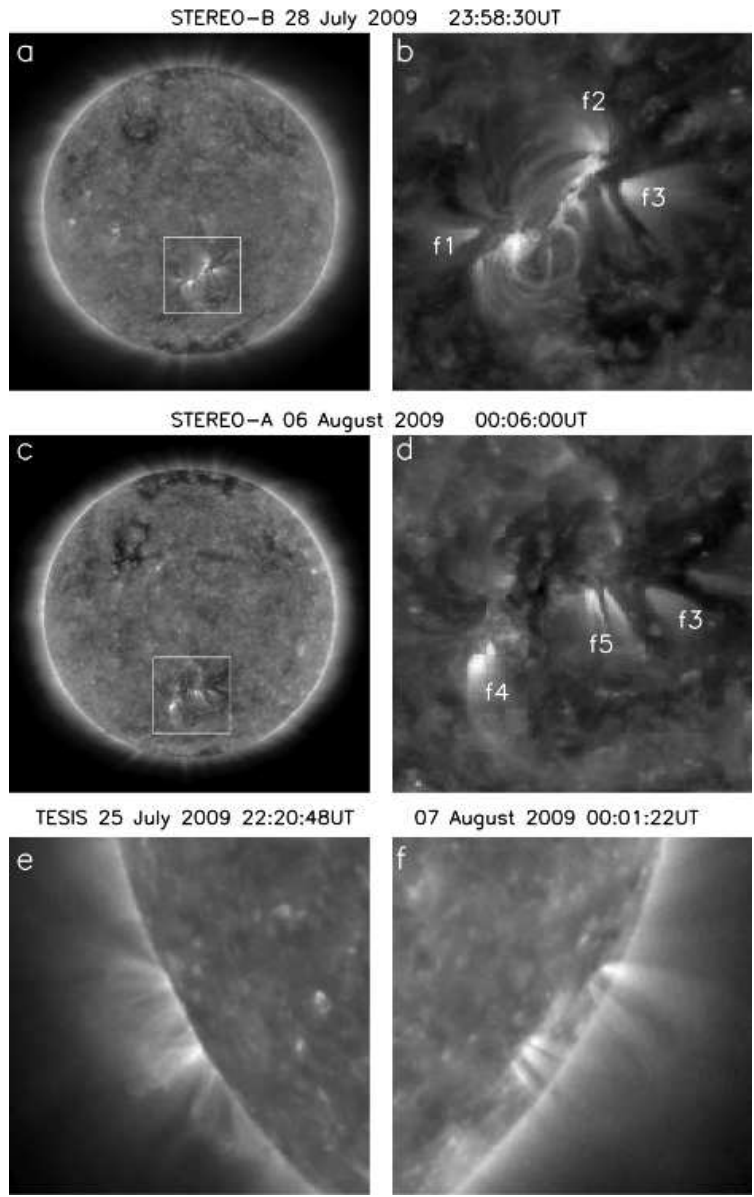}
              }
              \caption{Images of the AR in the 171 \AA\ band: at the disk taken by \textit{STEREO-B}/EUVI on July 28, 2009 (a) with zoomed box around the AR (b) and by {\it STEREO-A}/EUVI on August 6, 2009 (c) with zoomed box (d); at the eastern limb taken by   \textit{CORONAS-Photon}/TESIS on July 25, 2009 (e) and at the western limb on August 7, 2009 (f). Letters f1-f5 indicate the groups of fan loops.
                      }
   \label{fig2}
   \end{figure}

We investigate the relationship between outflows at the Sun and their coronal signatures by studying an isolated AR observed at the solar disk from July 25 to August 8, 2009. The AR, not numbered by NOAA (the former AR 1024 in the previous rotation), was observed near the central meridian on July 28 with the {\it STEREO-B} EUVI telescope \cite{Howard08} in the 171 \AA\ band (see Figure \ref{fig2} (a) and (b)). The structure of the AR consisted of the system of loops extended in the direction from South-East to North-West. Three groups of the fan loops f1--f3 were observed in the eastern and western sides of the AR. The fans have lower curvature and wider angular spread than the ordinary closed loops in the center of the AR.

The EUVI telescope on {\it STEREO-A} observed the AR near the central meridian nine days later, on August 6 (Figure \ref{fig2} (c) and (d)). During the period from July 28 to August 6 the AR structure has appreciably changed. In particular, the closed loops seen on July 28 in the AR center fully disappeared. The fan loops were also evolved: the f1 and f2 groups vanished, the brightness of f3 decreased, the new groups f4 and f5 appeared in the AR center, probably, in the roots of the former closed loops.

On July 25 the AR was observed at the eastern limb by the TESIS telescope onboard the CORONAS-Photon satellite. Figure~\ref{fig2} (e) and (f) displays two TESIS images taken on July 25 and on August 7, when the AR was located at the eastern limb and near the western limb, respectively. These TESIS images were the closest in time to the above mentioned {\it STEREO-B} and {\it STEREO-A} images (between July 25 and August 1, 11:26:40UT TESIS was out of operation). The coronal structure above the AR at the eastern limb contained mainly closed loops which agreed well with the configuration seen on the \textit{STEREO-B} disk image in Figure~\ref{fig2}(a). To August 7 the central closed loops transformed into extended low-curvature structures which may be the lower parts of high loops or coronal rays. Evolution of the AR can be seen in the TESIS movie in the Supplement to the on-line version of the article.

\begin{figure}    
   \centerline{\includegraphics[width=1.\textwidth,clip=]{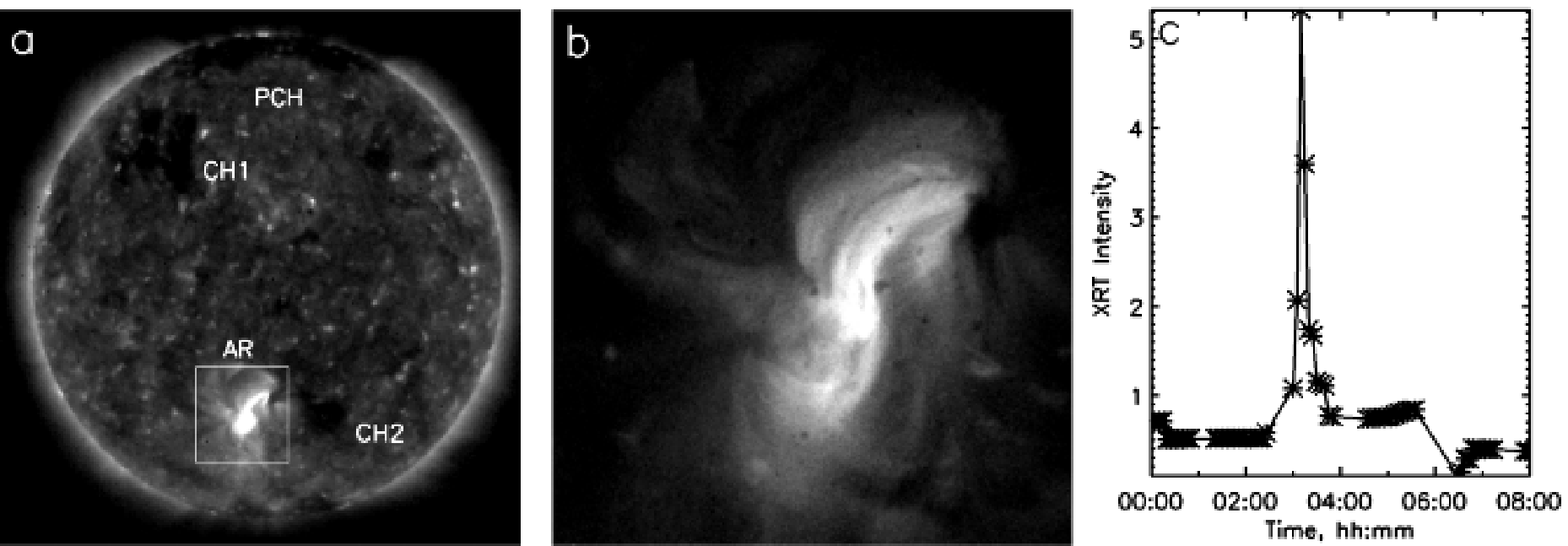}
              }
              \caption{(a){\it HINODE}/XRT full frame image  taken on August 1, 2009 at 06:21:03UT. The indicated objects are: the AR under study (AR), polar coronal hole (PCH) and two mid-latitude coronal holes (CH1 and CH2). The zoomed area is outlined by a box.  (b) One of the zoomed partial images of the AR taken at 01:23:44UT. (c) Light curve of the total X-ray flux integrated over the partial frame as a function of time from 00:00 to 08:00UT showing the flare peaked at 03:02UT.
                      }
   \label{fig3}
   \end{figure}

The X-ray and EUV instruments {\it Hinode}/XRT  and EIS, {\it SOHO}/EIT \cite{Delaboudiniere95} and {\it CORONAS-Photon}$/$TESIS \cite{Kuzin09} observed the AR at the disk close to the central meridian on August 1. Figure \ref{fig3}(a) shows the image of the Sun taken by the {\it Hinode}/XRT on August 1, 06:21:03UT, in the full frame mode. The zoomed image of the box in the full frame image is displayed in Figure \ref{fig3}(b). The main solar structures seen in the XRT image are the AR, the polar coronal hole (PCH), the mid-latitude coronal hole northward from the AR (CH1) and a small coronal hole adjacent to the western side of the AR (CH2).

A sequence of the partial frame images taken by XRT between 00:00 and 08:00 showed a flare peaking at 03:02 UT accompanied with a loop expansion. The light curve of the XRT total intensity in the partial frame as a function of time is shown in Figure \ref{fig3}(c). The {\it Hinode}/EIS instrument scanned the AR in the direction from right to left in a range of emission lines, from 01:22:14 to 02:57:14 UT. The flare started when the raster reached the left side of the frame, hence, the EIS scanned image corresponds to the undisturbed AR structure.

\begin{figure}    
   \centerline{\includegraphics[width=1.0\textwidth,clip=]{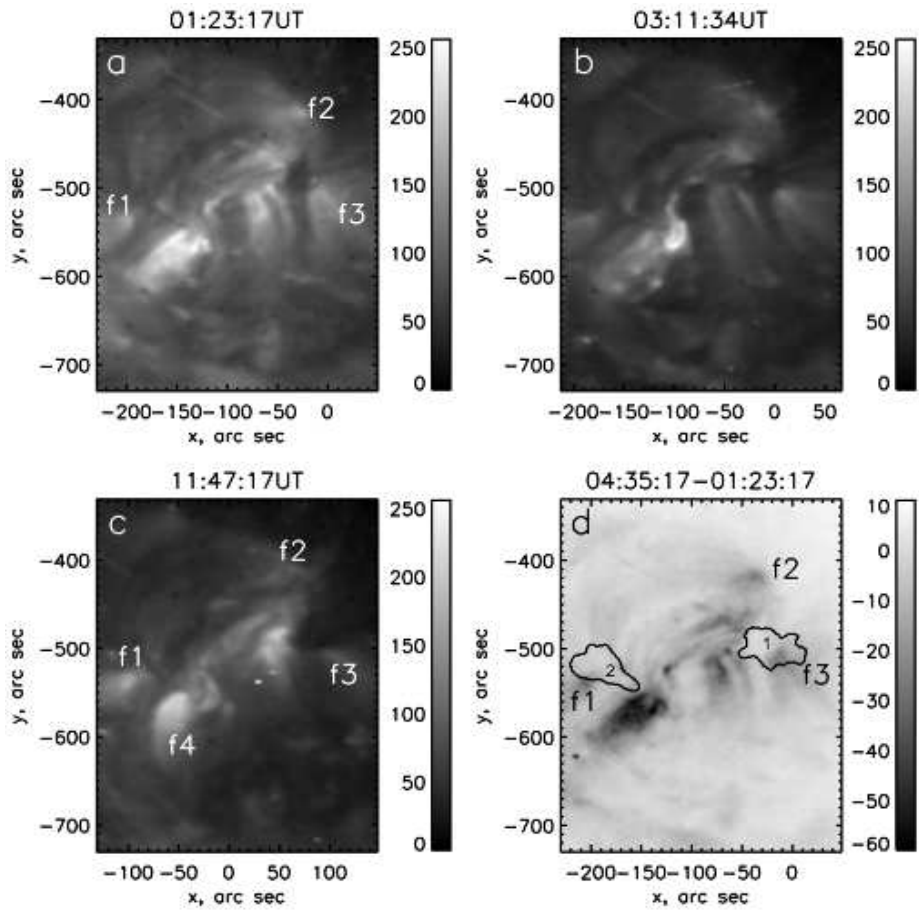}
              }
              \caption{The EIT images in 195 \AA\ taken on August 1, 2009 at the beginning of the EIS scan (01:23:17 UT,(a)); after the maximum of the flare (03:11:34 UT; (b)) and near the beginning of the TESIS observations (11:47:17 UT, (c)); the EIT difference image showing a dimming after the flare at 04:35:17 UT (d). The contoured regions 1 and 2 correspond to the largest areas of outflows with velocity $|V|>10$~\kms determined by EIS from the Doppler shift in the Fe~{\sc xii} emission line (see the next section). Letters f1--f4 indicate the fan loops in accordance with Figure \ref{fig2}.
                      }
   \label{fig4}
   \end{figure}

On August 1 EIT observed the solar disk in the course of its ordinary 12 min synoptic sequence in the 195 \AA\ band. The EIT images are very useful to make a bridge between the XRT and EIS observations in the beginning of August 1 and the following TESIS observations which started at 11:26:40 UT on August 1 and continued during a week when the AR reached the western limb. Between August 4 and 12 SOHO was in keyhole, so TESIS was the only one operating wide field EUV telescope which observed the corona above the AR at the western limb. Figure \ref{fig4} shows a series of EIT images in 195 \AA\ taken at the beginning of the EIS scan (01:23:17 UT, panel (a)), after maximum of the flare (03:11:34 UT, panel (b)) and near the beginning of the TESIS observations on that day (11:47:17 UT, panel (c)). The panel (d) shows the difference between the images taken at 04:35:17 UT and at 01:23:17 UT with the dimming appeared after the flare in the eastern part of the AR. At 01:23:17UT (Figure \ref{fig4}(a)) the positions and brightnesses of the fan loops f1--f3 were practically the same as seen by \textit{STEREO-B} on July 28 (Figure \ref{fig2}(b)). After the flare, at 11:47:17UT (Figure \ref{fig4}(c)) the f1 fan was unchanged, the brightnesses of f2 and f3 decreased by $\sim$10\%\ and a new fan group f4 appeared at the place of the former loops and the dimming seen in Figure \ref{fig4}(d). For further discussion the contours of the outflow regions determined by the EIS in the next section are superimposed on the difference image.

\subsection{EIS observations of outflows}

We analyzed the EIS image of the AR on August 1, 2009. The EIS instrument observes in two wavebands: 170--210 \AA\ and 270--290 \AA. It has four slit/slot positions: 1\arcs, 2\arcs, 40\arcs, and 266\arcs.  In this paper we will analyze data taken with the 2\arcs~ slit and using the fine mirror movement to ``raster'' and build up an image. The standard calibration was used and the emission lines were fitted with a Gaussian profile in order to measure the intensities and Doppler shifts. As a reference area for velocity calibration, we selected the quiet region in the right bottom of the EIS frame, where the mean velocity was set to zero.

\begin{figure}
\centerline{\includegraphics[width=1.0\textwidth,clip=]{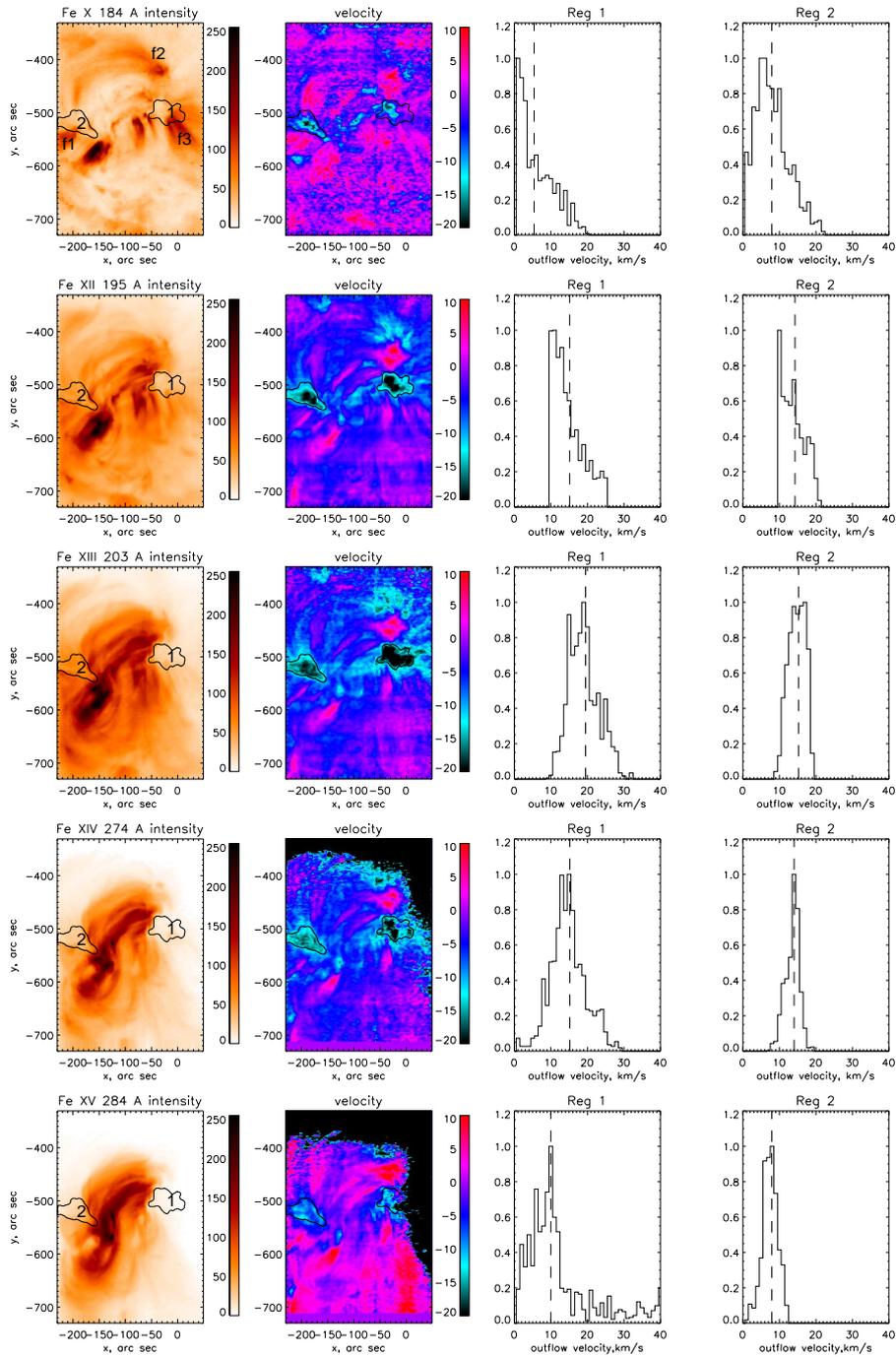}}
            \caption{The Fe~{\sc x}, Fe~{\sc xii}, Fe~{\sc xiii}, Fe~{\sc xiv} and Fe~{\sc xv} intensity and Doppler velocity maps for the 1 August 2009 EIS raster. The blue colour shows the upflowing plasma, and the red colour corresponds to the downflowing plasma. The contours correspond to the largest outflow regions defined from the Fe~{\sc xii} velocity map by the condition $|V| >$~10~\kms. The two right columns demonstrate histograms of the outflow velocity distributions in the regions 1 and 2 for corresponding Fe ions (dashed lines show the mean velocities). Letters f1--f3 indicate the fan loops pointed out in Figure~\ref{fig4}.}
\label{fig5}
\end{figure}

The intensity and the Doppler line-of-sight velocity maps are shown in Figure~\ref{fig5} for the Fe~{\sc x} 184.54 \AA\, Fe~{\sc xii} 195.12 \AA\ , Fe~{\sc xiii} 203.8 \AA\ , Fe~{\sc xiv} 274.2 \AA\ and Fe~{\sc xv} 284.16 \AA\ lines. The velocity maps contain both regions of outflows corresponding to the blue Doppler shift and downflows corresponding to the red shift.

In all ionic lines the outflows reach the largest velocities in two compact regions marked by numbers ``1'' and ``2'' in Figure~\ref{fig5}. Region 1 is located at the western boundary of the AR in the interface region adjoined to the coronal hole CH2. Region 2 is located in the eastern side of the AR in the vicinity of bright loops. For comparative analysis we contoured the areas of regions 1 and 2, where the outflow velocities in the strongest Fe~{\sc xii} line 195.12 \AA\ exceed 10 \kms, and superimposed them on all intensity and velocity maps. In all ionic lines the highest outflow velocities are concentrated in the central parts of these areas, whereas the line intensities decrease as the absolute value of velocity growing. The histograms of distributions of the outflow velocities integrated over the contoured areas of regions 1 and 2 are shown in Figure~\ref{fig5} in the two right columns. In both regions the mean outflow velocity (marked by dashed line) increases from Fe~{\sc x} to Fe~{\sc xiii} and then decreases for Fe~{\sc xiv} and Fe~{\sc xv}. Although the signal statistics for the Fe~{\sc xiv} and Fe~{\sc xv} ions in the region 2 is lower than that in the region 1, the principal dependence of the mean velocity on the ionization state is the same for both regions. This result differs from the conclusion of \inlinecite{DelZanna08} that the outflow velocities in ARs are larger for higher temperature lines.

In all lines the largest velocity does not exceed 30 \kms\ in projection on the line-of-sight except for the Fe~{\sc xv} line, in which the distribution for region 1 has a tail at higher velocity. Possibly, this tail can be produced by the minor high-speed component of outflow similar to that described by \inlinecite{Tian12}. The largest dimensions of regions 1 and 2 in the Fe~{\sc xii} line amount 45 and 58 Mm respectively, their variation with the ion number approximately follows the mean velocity dependence. The fan loops  f1--f3 introduced in Figures ~\ref{fig2}(b) and ~\ref{fig4}(a) are most clearly seen in the Fe~{\sc x} emission line, gradually vanishing in the lines of higher charged ions.

It should be noted that region 1 crossed the central meridian between July 31, 23:31UT and August 1, 08:46UT, region 2 -- between August 1, 20:23UT and August 2, 06:47UT, that is considerably differ from the times of the EIS scan. We did not scan the AR with EIS after the flare and only suppose that region 2 remained after this evolution to the beginning of the TESIS observations because in the EIT images in Figure~\ref{fig4} the brightness and position of the fan loops f1 at 11:47:17UT did not changed in comparison with that at 01:23:17UT.

       \begin{figure}    
   \centerline{\includegraphics[width=1.0\textwidth,clip=]{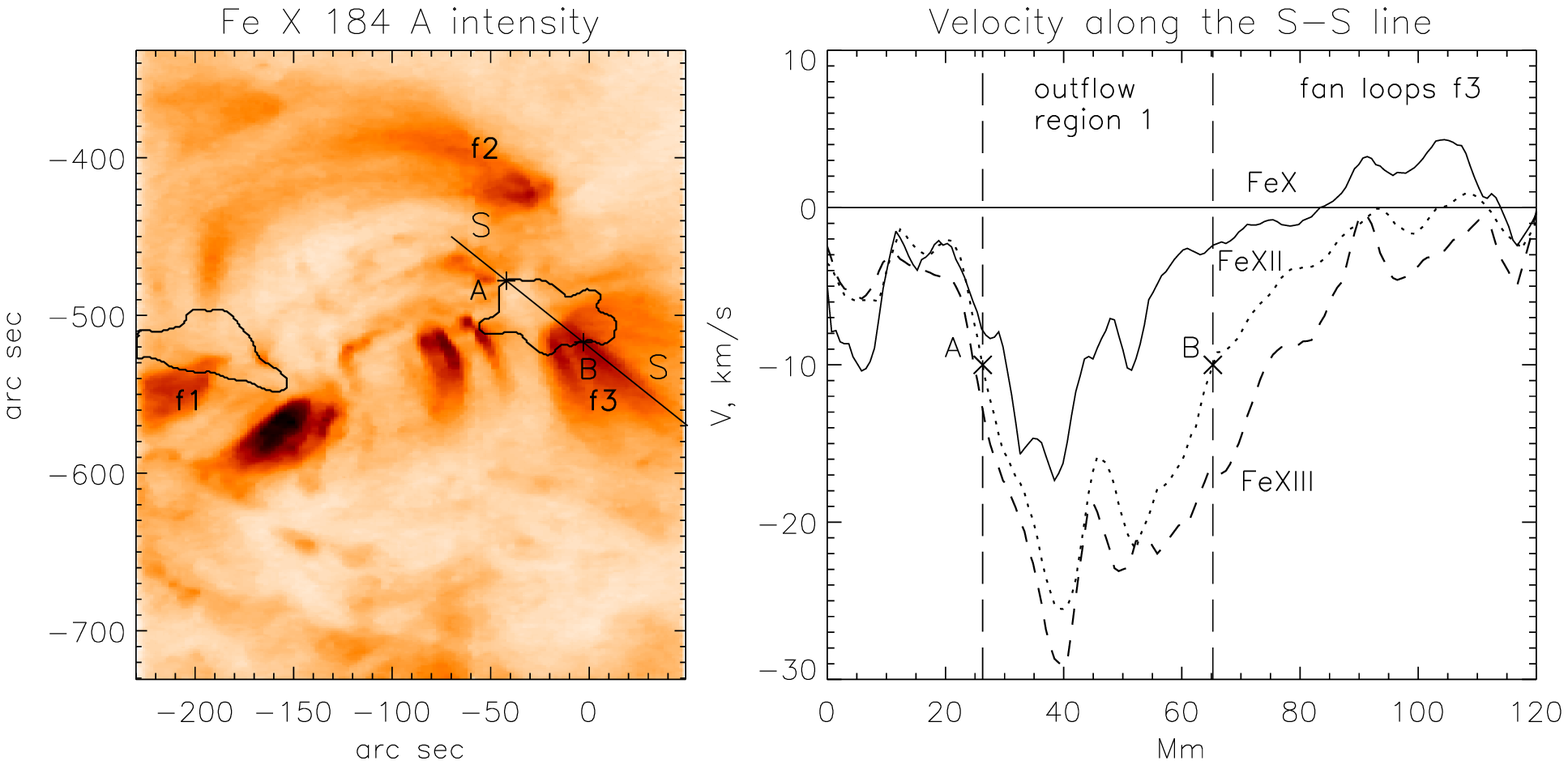}
              }
              \caption{Left panel: position of the ``S-S'' scan line crossing region 1 and adjacent fan loops in the Fe~{\sc x} intensity map. The right panel:  the Doppler velocity as a function of length (in Mm) along the scan direction for different Fe-ions: Fe~{\sc x} (solid line), Fe~{\sc xii} (dotted line) and Fe~{\sc xiii} (dashed line). Letters ``A'' and ``B'' indicate the boundaries of region 1. ``f1'', ``f2'' and ``f3'' designate the fan loops according to Figure~\ref{fig4}
                }
   \label{fig6}
   \end{figure}

To understand a relationship between fan loops and plasma outflows we scanned the velocity maps across the outflow region 1 and along the fans in different Fe-ion lines. Figure~\ref{fig6} shows the ``S-S'' scanning line which crossed region 1 and adjacent fan loops f3. The velocity in the Fe~{\sc x} spectral line inside region 1 is negative reaching the maximum of $|V|$ = 17 \kms\ in its center. In the fan loops f3 the velocity becomes positive indicating downflow with maximal value of 5 \kms\ and then falls down to zero. In the Fe~{\sc xii} and Fe~{\sc xiii} lines the velocities are negative along the whole scan length reaching in the center of outflow the values of -25 and -30~\kms\ correspondingly and gradually decreasing to about zero in the fan loops (within the typical EIS accuracy of $\pm$ 2--3 \kms\ ). It was checked that this relationship was the same for the fans f1 and f2.

Using the EIS intensity and velosity maps, we estimated the mass flux balance within the EIS frame in various Fe-ion lines. We define the mass flux density as \textit{n$\ast$v} , where \textit{n} and \textit{v} are local values of the plasma density and velocity. As a proxy of density \textit{n}, we used the square root of the line intensity taking into account that the Fe-ion lines in the lower corona are excited by the electron-ion collisions with assumption that the temperature and the ion abundances are constant. Earlier similar approach was used by \inlinecite{Marsch04}. Integration of the bidirectional flux over the EIS FOV has shown that the total outflux in the Fe~{\sc x} line was near in balance with the total downflux: the outfux/downflux ratio is $\sim$~ 1.4. In the Fe~{\sc xii} line this ratio increased to $\sim$~7, in the Fe~{\sc xiii}--Fe~{\sc xiv}  to 20--30. It means that in all lines the resulting total flux from the AR was directed outwards at least within the boundaries of the EIS frame.

\subsection{DEM analysis of plasma temperature distributions}

To explore the properties of the outflowing plasma in comparison with other coronal regions, we applied the DEM analysis to different areas in the AR. For that we used the diagnostic technique developed within the framework of the probabilistic approach to the spectral inverse problem for determining the temperature content of the emitting plasma sources (see \opencite{Urnov07}; \opencite{Goryaev10}). This inversion technique called Bayesian iterative method (BIM) based on the Bayes' theorem is used to reconstruct DEM distributions with the ultimate resolution enhancement (super-resolution) as compared with linear and other non-parametric methods (see \opencite{Gelfgat93} for details, and references therein). The latter property allows the BIM to reconstruct fine temperature structures in the DEM profiles where other DEM techniques derive smoothed and less informative solutions.

\begin{figure}    
   \centerline{\includegraphics[width=1.0\textwidth,clip=]{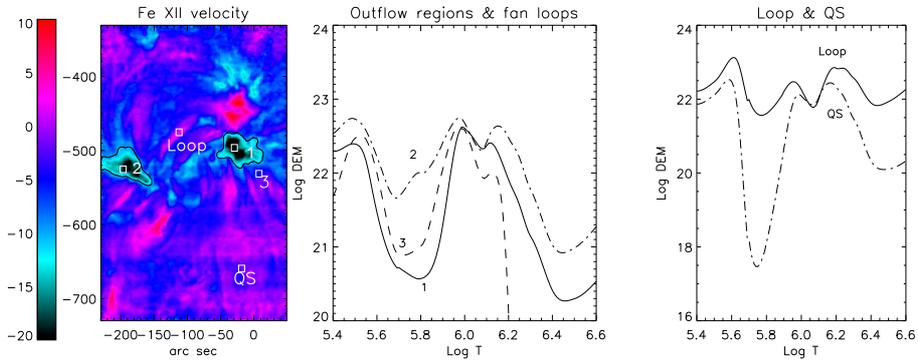}
              }
              \caption{The DEM functions for different solar structures indicated by boxes on the Fe {\sc xii} velocity map (left panel): 1 and 2 -- outflow regions, 3 -- fan loops f3, Loop -- closed loops in the center of the AR, QS -- quiet Sun.
                }
   \label{fig7}
   \end{figure}

 The DEM analysis was applied to the five areas in the AR which are displayed as boxes of 10$\times$10 pixels on the Fe~{\sc xii} velocity map in Figure~\ref{fig7} (left panel). The measured intensity fluxes in EIS spectral lines were obtained using the standard EIS solarsoft. The corresponding EIS lines are listed in Table \ref{T-simple1}. In this table $T_{\mathrm{max}}$ is the temperature of the maximum abundance for each ion \cite{Mazzotta98}. To calculate the contribution functions we used the CHIANTI database (version 6.0.1 -- \opencite{Dere97}; \opencite{Landi06}). The DEM analysis was performed adopting the ion fractional abundances of \inlinecite{Mazzotta98} and the coronal elemental abundances of \inlinecite{Feldman92}.

In order to calculate the contribution functions, we also carried out electron density measurements using line ratios. There are a number of density-sensitive line ratios in the EIS wavelength ranges (see, e.g., \opencite{Young09}). For the density diagnostics we used the Fe {\sc xii} $\lambda$186.88/$\lambda$195.12 ratio sensitive in the range of densities $\log N_e\approx$~8--12 ($N_e$ in cm$^{-3}$). The evaluated values of densities for the solar areas under study are found to be $\log N_e\approx$~8.4--9.

\begin{table}
\caption{List of the EIS lines used in the DEM analysis.
}
\label{T-simple1}
\begin{tabular}{lcc|lcc}     
  \hline                   
Ion & $\lambda$ , \AA\ & $\log T_{max}$ & Ion & $\lambda$ , \AA\ & $\log T_{max}$ \\
  \hline
Fe {\sc x} & 184.54 & 6.0 &  Fe {\sc xiii} & 203.83 & 6.2   \\
Fe {\sc xii} & 186.88 & 6.1 &  Fe {\sc xiii} & 203.80 & 6.2  \\
Fe {\sc xi}& 188.23 & 6.1 & Fe {\sc xii} & 203.72 & 6.1     \\
Fe {\sc xi} & 188.30 & 6.1 & Fe {\sc viii} & 186.60 & 6.0   \\
Fe {\sc xii} & 192.39 & 6.1 & Fe {\sc x}& 257.26 & 6.0    \\
Fe {\sc xi} & 192.83 & 6.1 & Fe {\sc xiv} & 264.79 & 6.2    \\
O {\sc v} & 192.90 & 5.4 & Fe {\sc xiv} & 274.20 & 6.2    \\
Fe {\sc xii} & 195.12 & 6.1 & Si {\sc vii} & 275.35 & 5.8    \\
Fe {\sc xii} & 195.18 & 6.1 &  Fe {\sc xv} & 284.16 & 6.3   \\
Fe {\sc xiii} & 202.04 & 6.2 &  Fe {\sc xiv} & 257.39 & 6.3   \\
  \hline
\end{tabular}
\end{table}

Figure~\ref{fig7} shows results of the DEM analysis for the solar areas marked by boxes: two outflow regions and fan loops near region 1 (middle panel), closed loop and quiet solar regions (right panel). It is seen that all DEM distributions contain three prominent peaks which are related to corresponding temperature components of plasma overlapped along the line-of-sight at $\log T\sim$~5.5--5.6, $\log T \sim$~5.9--6.0, and $\log T\sim$~6.1--6.2. All distributions have low temperature peak at temperature of $\log T\approx$~5.5--5.6 ($T\sim$~0.4 MK), which produces emission of  colder lines of O {\sc v} and Si{\sc vii} in the transition region along the line of sight. The high temperature peak at $\log T\approx$~6.2 corresponding to closed loops has significantly larger emission measure than the peaks at temperatures of $\log T\approx$~6.0. The DEM distribution for the QS region demonstrates that the high temperature peak at $\log T\approx$~6.1--6.2 exceeds the middle peak at $\log T\approx$~5.9--6.0.

In both outflow regions (middle panel in Figure~\ref{fig7}) the situation is contrary to the QS and loop DEM structures. The medium temperature peak at $\log T \approx$~6.0 is larger than the high temperature peak indicating that the plasma component with $T\approx$~1 MK dominates in the outflow regions. The high temperature component with $\log T \approx$~6.2 ($T\approx$~1.5 MK) is less discernible than in closed hot loop structures. However, the hotter component is essentially weaker in the fan loops than in the outflow regions. The inferences from our DEM analysis are in particular consistent with the results obtained by \inlinecite{Brooks2011} who used the Markov Chain Monte Carlo (MCMC) algorithm for the reconstruction of the DEM distributions. Brooks and Warren did not provide the calculated DEM profiles, but the temperatures of the DEM peaks found by these authors for the outflow regions agree with the temperatures obtained in the present work.

\subsection{Magnetic field in the AR and its relationship with coronal rays}

\begin{figure}    
   \centerline{\includegraphics[width=1.0\textwidth,clip=]{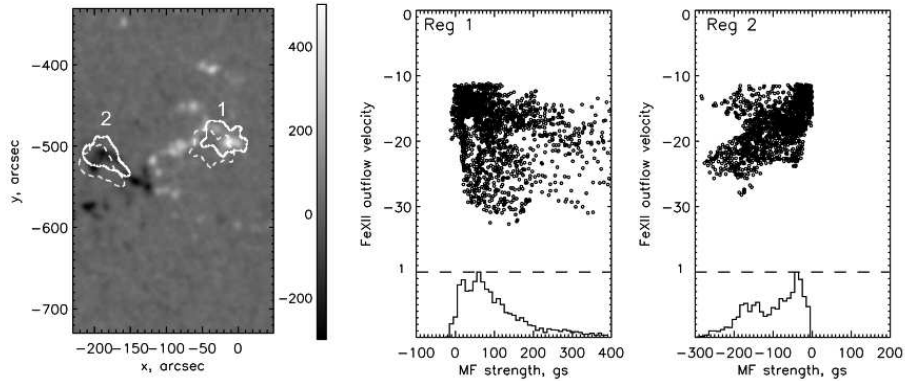}
              }
              \caption{Line-of-sight photospheric magnetic field map from the Big Bear observatory data (left panel), and  distribution of outflow radial velocities in dependence on magnetic field strength in regions 1 (middle) and 2 (right). The dashed contours on the map correspond to the outflow regions in the Fe {\sc xii} line measured by EIS, solid contours -- to their positions projected  to the photosphere. The graphs in the bottom show normalized distributions of the summarized radial velocities on magnetic field strength.
                      }
   \label{fig8}
   \end{figure}

The left panel of Figure~\ref{fig8} displays the map of photospheric magnetic field within the EIS FOV taken from the data obtained at the Big Bear observatory on August 1, 2009 at 01:22UT. The magnetic field in the AR area had a simple dipole configuration being negative at the eastern side and positive at the western side. The contours superimposed on the map outline the boundaries of the outflow regions 1 and 2 determined from the EIS Fe~{\sc xii} velocity map: the dashed contours show their measured positions as in Figure~\ref{fig5}, the solid contours -- their calculated positions at the photosphere after correction of the projection effect in a simple geometrical model. We assume that the outflowing plasma near the solar surface propagates along the open magnetic field lines directed normally to the surface and seen by the observer at the mean latitudinal angle of 32$^{0}$ to the South. We compensated the projection effect assuming that the Fe~{\sc xii} line is emitted from the height $h\approx$~25~Mm which is a half of the density hydrostatic scale of height $H_{\mathrm{hst}}$=50 Mm for $T\sim$~1~MK (due to collisional excitation of this line, its intensity is proportional to the density squared, so the intensity scale of height should be equal to 1/2 of its value for density). Then we calculated the radial velocities from their measured line-of-sight values by division on cosine of the angle between the local normal and the line-of-sight.

The middle and right panels in Figure~\ref{fig8} display correlation between the radial outflow velocities and the line-of-sight magnetic field strength in regions 1 and 2 projected to the photosphere. The histograms show distributions of summarized radial velocity versus magnetic field strength. The results show that: (1) outflows emerge from regions of both magnetic field polarities, and (2) the bulks of outflows in both regions  with $|V|>$~10~\kms\ are concentrated at the magnetic fields $|B|\leq$~200~G aside of the regions of the strongest field ($|B|_\mathrm{max}\approx$~300--400~G) in the AR. These results differ from the case of plasma flows along closed loops studied by \inlinecite{Marsch04}. They found that the outward mass flows corresponded to the strongest fields more than 200 G in the region of negative field polarity while the inward mass flows were associated with positive polarity.

    \begin{figure}    
   \centerline{\includegraphics[width=0.9\textwidth,clip=]{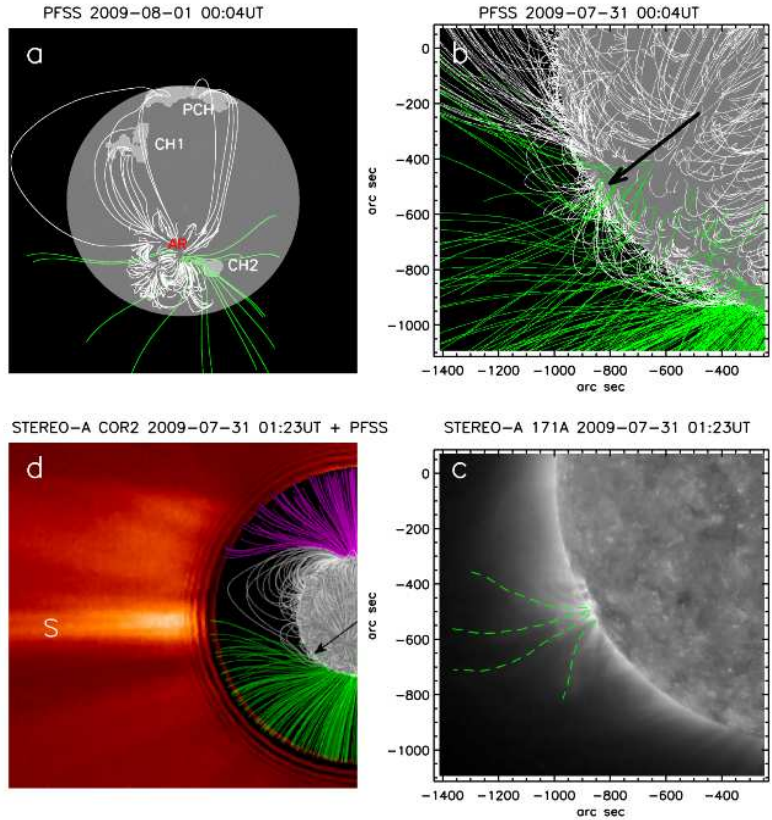}
              }
              \caption{(a) Magnetic field connections of the AR with other structures at the disk derived with PFSS for August 1, 00:04UT: PCH - the polar coronal hole, CH1 and CH2 - the mid-latitude coronal holes. (b) The partial image of the PFSS magnetic field map at the eastern limb calculated for July 31, 00:04UT seen from the position of \textit{STEREO-A} at the heliographic longitude 320$^0$ and latitude 7$^0$. (c) The \textit{STEREO-A} EUVI partial image of the corona in the 171~\AA\  band taken on July 31 at 01:23UT. The dashed green lines trace the brightest coronal rays from the ARCH in the inner corona. (d) The WL image of the corona obtained by \textit{STEREO-A} COR2 on July 31 at 01:23UT combined with the 3D map of open magnetic field lines seen from the same viewing point. The open field lines connect the ARCH with the equatorial streamer S. The arrows point to the ARCH location. Closed magnetic field lines are marked by white, open positive/negative lines - by green/magenta colors.}

   \label{fig9}
   \end{figure}

        \begin{figure}    
   \centerline{\includegraphics[width=0.9\textwidth,clip=]{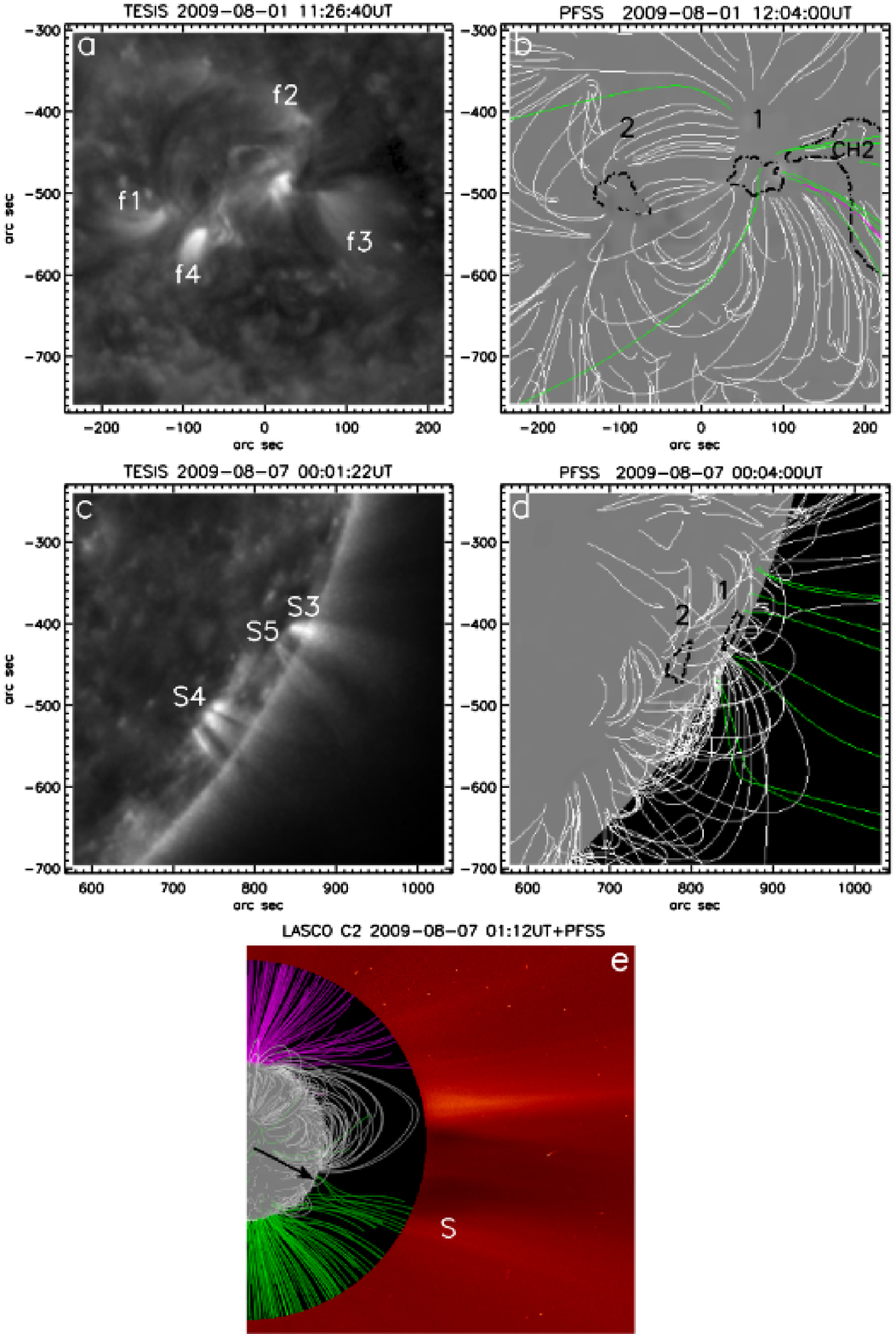}
              }
              \caption{The zoomed images of the AR taken by TESIS in the 171 \AA\ band at the disk center on August 1 at 11:26:40UT (a) and at the western limb on August 7 at 00:01UT (c). Magnetic field maps calculated with the PFSS extrapolation for August 1, 12:04UT (b), and August 7, 00:04UT (d). The WL image of the corona taken by LASCO C2 on August 7 at 01:12UT combined with the PFSS 3D map for 00:04UT (e). The closed and open magnetic field lines are marked by colors as in Figure~\ref{fig9}. The dashed contours outline positions of the outflow regions projected to the photosphere and differentially rotated to the time of the PFSS extrapolation. The scales are in arc seconds relative to the solar center. The coronal structures S4, S5 and S3 correspond to the appropriate fans in Figure~\ref{fig2}(d). The arrow indicates the position of the ARCH at the limb which is connected by open field lines with the streamer S.
                      }
   \label{fig10}
   \end{figure}

The global magnetic connections of the AR with other solar regions on August 1, 2009, 00:04UT (the closest in time to the EIS scan) calculated with the Solarsoft PFSS code \textit{pfss\_viewer} are shown in Figure~\ref{fig9}(a).  There are seen three groups of field lines: open field lines with positive polarity in the western part of the AR neighboring to the coronal hole CH2 (hereafter we will name this region as the ARCH), short loops connecting the positive and negative magnetic plagues in the AR and long loops connecting the AR with the northern PCH and CH1. Closed loops are marked by white color, open positive/negative lines -- by green/magenta colors. Figure~\ref{fig9}(b) displays the magnetic field configuration calculated for July 31 2009, 00:04UT as it seen from the position of \textit{STEREO-A} at heliographic longitude 320$^0$ and latitude 7$^0$. The corresponding image of the corona taken in the \textit{STEREO-A} EUVI 171 \AA\ band on July 31, 2009, at 01:23UT is presented in Figure~\ref{fig9}(c). This time was selected due to better visibility of the coronal rays above the ARCH expanding along open magnetic field lines (three longest rays are marked out by dashed green lines). Figure~\ref{fig9}(d) gives a view of the WL corona provided by \textit{STEREO-A} COR2 on July 31, 2009 at 01:23UT combined with the magnetic field structure seen from the same viewing point. These pictures demonstrate that one day before EIS detected outflows at the disk, \textit{STEREO-A} observed coronal rays co-aligned with open field lines connecting the ARCH with the stalk of the equatorial streamer S.

In Figure~\ref{fig10} the zoomed image of the ARCH taken by TESIS in the 171 \AA\ band on August 1, 11:26:40UT (a) is compared with corresponding PFSS magnetic field map calculated for 12:04UT (b).  Superimposed are the contours of the outflows 1 and 2 from the EIS velocity map (projected to the photosphere) together with the contour of the CH2 coronal hole, all rotated to the PFSS mapping time. The structure of the AR and positions of the fan loops f1--f4 in the TESIS image were similar to that shown in the EIT 195 \AA\ image in Figure~\ref{fig4}(c). The open field lines of positive polarity are originated nearby the position of the western outflow region 1. The fan loops f3 go out from this area in the southwestern direction along closed magnetic lines. The fan loops f2, most probably, correspond to the group of long magnetic loops connecting the AR with the polar coronal hole (PCH) shown in Figure~\ref{fig9} (a). The fan loops f1 and f4 are co-aligned with closed magnetic lines connecting plagues of opposite polarities in the AR.

Figure~\ref{fig10}(c) displays the TESIS image of the coronal structures above the AR taken a week after, on August 7, at 00:01:22UT, when the AR was positioned at the western limb and CH2 turned over the limb. According to Figure~\ref{fig2}(d), to this time the AR structure was seriously evolved in comparison with that displayed in Figure~\ref{fig10}(a) for August 1. The bright coronal structures S4, S5 and S3 at the limb were rooted approximately in the positions of the fan groups f4, f5 and f3 in Figure~\ref{fig2}(d). These structures had low curvatures which suggests that they may be lower parts of long loops or coronal rays. The magnetic field configuration presented in Figure~\ref{fig10}(d) consisted of closed loops connecting the structures S4 and S5 and open field lines in close vicinity to S3. The dashed contours superimposed on the magnetic field map outline positions of the regions 1 and 2 differentially rotated to the mapping time. Contour 1 coincides with the root of the coronal structure S3, whereas there are no any visible structures located near contour 2. In analogy with panel (b), if we assume that open field lines are rooted in region 1 as before, the structure S3 will be co-aligned with open field lines, so it can be identified as a bundle of coronal rays.  Figure~\ref{fig10}(e) displays the WL image of the corona at the western limb taken by LASCO C2 on August 7 at 01:12UT combined with the PFSS 3D map calculated for 00:04UT. The arrow indicates the position of the ARCH which is connected by open field lines with the weak streamer S inclined on $\sim22^0$ in the southwestern direction.

Summarizing, although spatial resolution of the EUV images as well as of the PFSS magnetic field maps at the limb is insufficient for direct coincidence of the coronal structures with open field lines, a topology of the calculated magnetic field suggests that the coronal rays observed on August 1 and August 7 represent signatures of outflows from the ARCH propagating along open field lines to the heliosphere.

\subsection{Identification of outflows with the solar wind sources derived by the WSA model}

\begin{figure}    
   \centerline{\includegraphics[width=1.\textwidth,clip=]{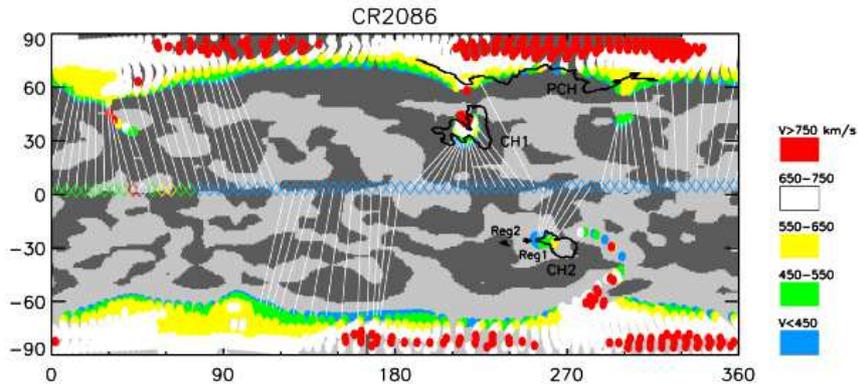}
              }
              \caption{The WSA map of CR 2086 (June 23 - August 19, 2009). White and grey background indicate areas of positive and negative photospheric magnetic field. The colored dots  represent the footpoints of open field lines coded according to the associated magnetic field expansion factors and predicted solar wind speeds in the ecliptic plane as shown on the scale at the right. White lines connect the supposed solar wind sources with the longitudes at which the corresponding open field lines reach the sub-earth points at the SS (marked by diamonds). Reg 1 and Reg 2 indicate positions of the EIS outflow regions transformed to the heliographic coordinates, the black contours correspond to the boundaries of PCH, CH1 and CH2.
              }
   \label{fig11}
   \end{figure}

As it was mentioned in Introduction, the WSA model identifies the solar wind source regions based on their association with regions of open field lines (coronal holes) given by the potential field extrapolation. Figure~\ref{fig11} shows the WSA map of the derived coronal holes calculated for Carrington rotation 2086 using the Mount Wilson photospheric magnetic field data (by courtesy of Y.-M. Wang, the other version of this map was published in ~\opencite{Wang10}\footnote{Comment of Y.-M. Wang: the previous version of the WSA map was built using an average of the photospheric field maps from Wilcox and Mount Wilson Solar observatories.}). The spatial resolution of the map is $\sim5^0$. White and grey areas correspond to regions of positive and negative photospheric magnetic field. The dots indicate the footpoints of open field lines with different values of magnetic field expansion factor and predicted solar wind speeds in the ecliptic plane coded by colors according to the scale shown at the right. White lines connect the supposed solar wind sources with the diamonds pointing out the longitudes at which the open field lines reach the sub-earth points at the SS. The superimposed black spots marked as Reg 1 and Reg 2 indicate positions of the EIS outflow regions transformed to the heliographic coordinates, the black contours correspond to the boundaries of PCH, CH1 and CH2 shown in Figure~\ref{fig9}(a). According to this map, the solar wind at the SS between longitudes 185 and 255$^0$ may be produced by CH1 and partially by PCH, between the longitudes 255 and 285$^0$ - by the source region which well coincides with the position of the ARCH. There are no open field lines in the vicinity of region 2, therefore, according to the WSA model, it cannot be recognized as a solar wind source. It should be noted that in the previous version of the WSA map published in \inlinecite{Wang10} the both outflow regions were not identified as the solar wind sources which demonstrates that the results of the WSA model are strongly depend on quality of the initial photospheric magnetic maps.

\subsection{\textbf{Association of outflows with the solar wind measured near the Earth}}

To reveal possible relationship of outflows from the AR with the solar wind near the Earth, we analyzed the solar wind parameters measured by instruments on several satellites: by SWEPAM \cite{McComas98} and SWICS \cite{Gloeckler98} on {\it ACE}, SWE \cite{Ogilvie95} on {\it WIND}, both rotating around  the Lagrangian L1 point, PLASTIC \cite{Galvin08} on {\it STEREO-B} and {\it STEREO-A} at their orbital positions displaced on some increasing angles in the ecliptic plane from the Sun-Earth line. To compare the solar wind data with the WSA map we projected those to SS in a function of heliographic longitude using the simple ballistic model based on the Archimedian spiral approximation and the measured at L1 values of the solar wind speed. We used the following relation between temporal records of the data and longitudes corresponding to the specified spacecraft at the given time: $Lon_{sc}(t)= Lon_{0}+\varphi+(t_{0}-t+\Delta T_{b})\times\omega$, where $Lon_{0}$ is the reference solar longitude seen from the Earth at the time $t_{0}$, $t$ the time of data record in units of Day of the Year (DOY), $\varphi$ the angle in the equatorial plane between the spacecraft position and the Sun-Earth line, $\omega$ the angular speed of the solar rotation as seen from the spacecraft, and $\Delta T_{b} = L_{ss}/V_r(t)$ is the ballistic traveling time delay, where $L_{ss}$ is the radial distance between the solar surface at $2.5 R_{\mathrm{sun}}$  and the spacecraft, $V_r(t)$ the solar wind radial velocity measured at the spacecraft. For our calculation we used the reference solar longitude $Lon_{0}$ = 251.5 for the reference time $t_{0}= 213.0$ DOY as seen from the Earth on August 1, 2009 at 00:00UT. For \textit{ACE} and \textit{WIND} we used $\varphi$ = 0$^0$ and $\omega$ = 13.23 degrees/day, for \textit{STEREO-A}  $\varphi$ = 57.6$^0$ and $\omega$ = 13.13 degrees/day, for \textit{STEREO-B}  $\varphi$ = -49.9$^0$ and $\omega$ = 13.32 degrees/day. All these parameters as well as distances between the spacecrafts and the source surface correspond to the reference time.

    \begin{figure}    
   \centerline{\includegraphics[width=0.7\textwidth,clip=]{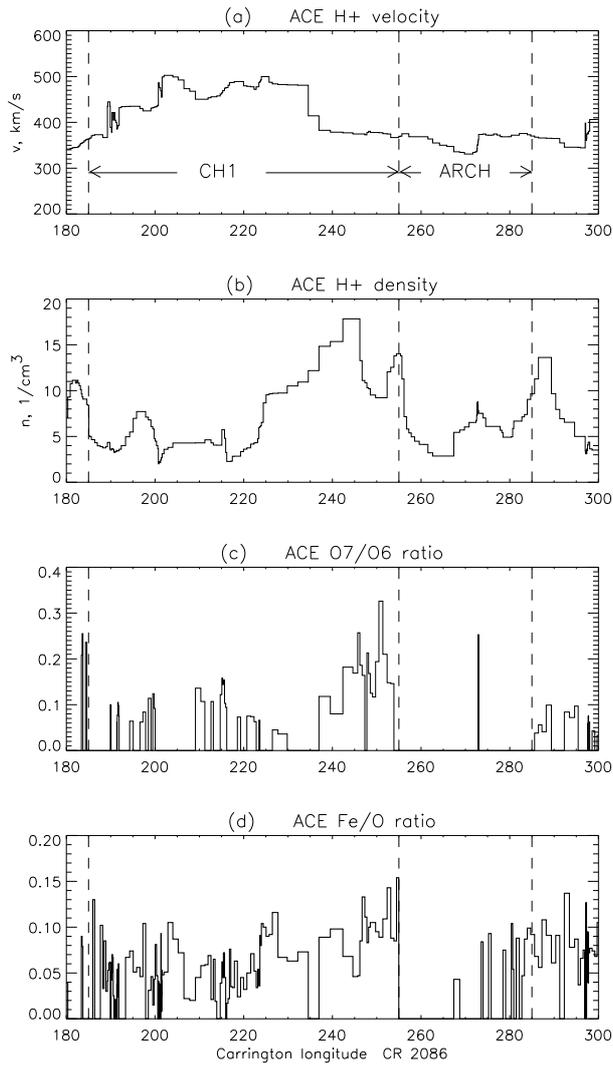}
              }
              \caption{The {\it ACE} solar wind velocity (a), density (b), ionic ratios of O$^{7+}$/O$^{6+}$ (c) and Fe/O (d) as a function of heliographic longitude at SS determined by the ballistic model.
                      }
   \label{fig12}
   \end{figure}

Based on the WSA map it was determined that the solar wind from the ARCH region was registered by \textit{ACE} and \textit{WIND} in the time interval between DOY 215.1 and 217.4 (August 3--5) with the mean travelling time delay $<\Delta T_{b}>$ = 4.8 days. The \textit{STEREO-B} spacecraft measured the solar wind from the same longitudinal range during DOY 211.8--213.5 (July 30--August 1) with $<\Delta T_{b}>$ = 5 days, \textit{STEREO-A} -- during DOY 220.2--221.9 (August 8--9) with similar time delay. Based on results of \inlinecite {MacNeice11} we suppose that the accuracy of our identification is $\sim$10$^0$ or 0.7 day.

Figure~\ref{fig12} shows the solar wind data measured by {\it ACE}:  velocity (a), density (b), O$^{7+}$/O$^{6+}$ ionic ratio (c) and Fe/O ratio (d) averaged over 1h intervals as a function of longitude at SS. It is seen that the ARCH region produced the slow wind with a velocity of $v$ = 320--380~\kms being well separated in time from the faster wind emerged from CH1 with $v$ = 400--500~\kms, so application of the simple ballistic method for the solar wind binding is fully justified. In spite of large gaps in the ACE O$^{7+}$/O$^{6+}$ and Fe/O data their mean values (averaged over non-zero data) are well correspond to the identified source regions. The mean O$^{7+}$/O$^{6+}$ ratio is 0.12 for the wind from CH1 and 0.25 (only one non-zero point) for the ARCH which agree with the values given by \inlinecite{Liewer04}: less than 0.2 for typical coronal hole and 0.2--0.3 for the ARCH interface region (the oxygen freezing-in temperature $T = 1.6$--$1.7$~MK). The mean Fe/O ratio is 0.06 for CH1 and 0.1 for the ARCH being in the limits 0.05--0.07 established for the fast wind from coronal holes and 0.09--0.13 for the wind from streamer belt at solar minimum \cite{Steiger00,Liewer04}.

        \begin{figure}    
   \centerline{\includegraphics[width=0.8\textwidth,clip=]{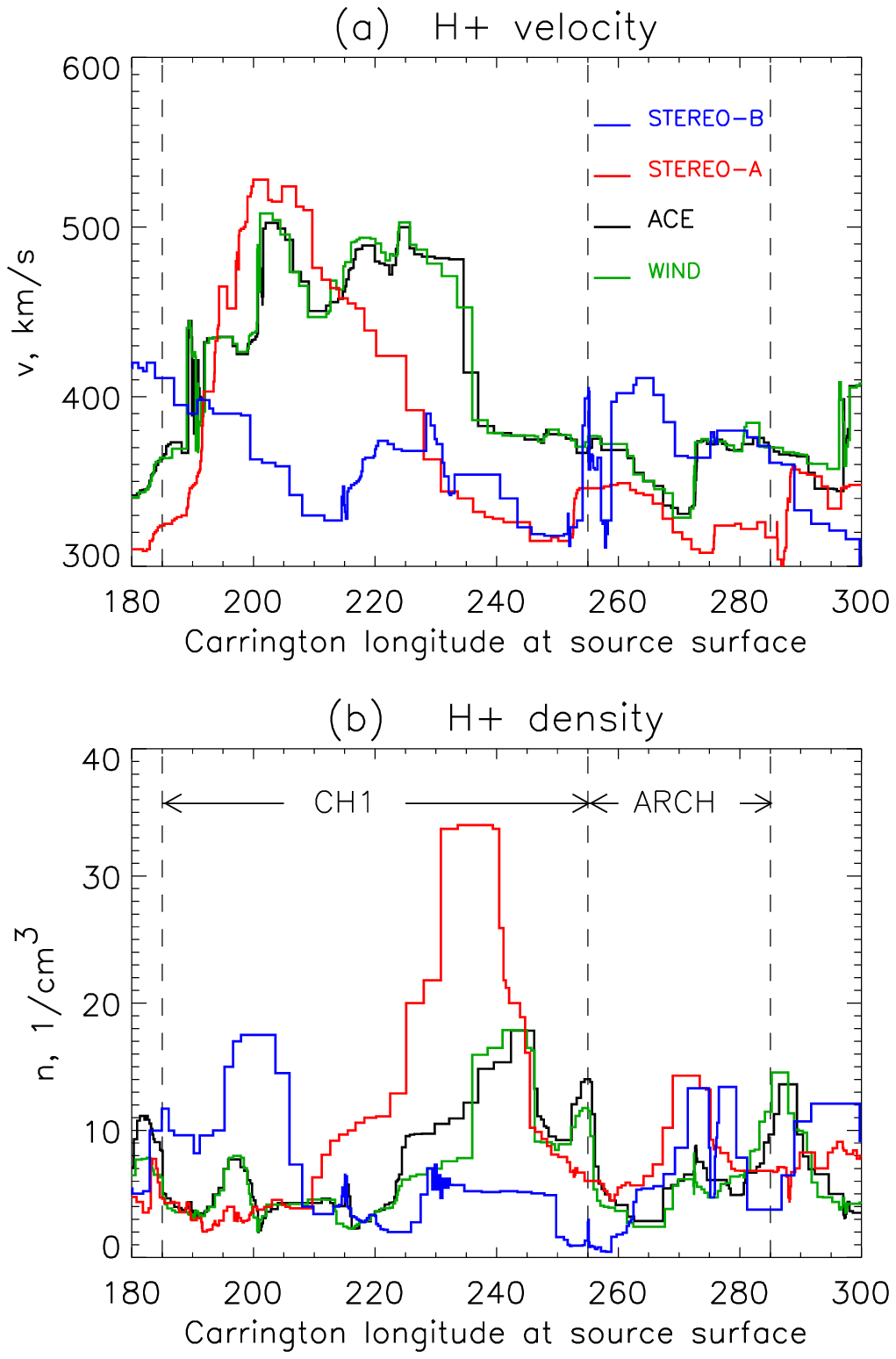}
              }
              \caption{The solar wind velocity (a) and  density (b) measured by {\it ACE} (black line), {\it WIND} (green line), {\it STEREO-B} (red line) and {\it STEREO-A} (blue line) as a function of heliographic longitudes at SS. The dashed lines delimit longitudinal ranges corresponding to the CH1 and the ARCH  source regions in accordance with the WSA map.
              }
   \label{fig13}
   \end{figure}

Figure~\ref{fig13} demonstrates a comparison of the solar wind parameters determined sequentially by \textit{STEREO-B}, \textit{ACE}, \textit{WIND} and \textit{STEREO-A} and presented as functions of longitude at SS. The longitude ranges associated with CH1 and the ARCH source regions are taken from the WSA map.  The panel (a) shows the solar wind (protons) velocity measured by four spacecrafts in three time intervals. In both latitudinal ranges associated with CH1 and the ARCH the \textit{ACE} and \textit{WIND} velocities fully coincide. The velocities measured by \textit{STEREO-B} and \textit{STEREO-A} considerably differ in both ranges due to temporal evolution of the CH1 and the ARCH structures. In all data the velocities associated with the ARCH source region are within the limits of 300--420~\kms. Many details in the velocity and density distributions for different spacecrafts  are well correlated in latitude which confirms adequacy of our time-longitude conversion within the accuracy of $\sim$10$^0$.

The density distributions shown in Figure~\ref{fig13}(b) have several enhancements in the longitudinal range of CH1 as well as in that of the ARCH. The density peak at longitudes 225--245$^0$ is most likely produced by the stream interaction. Its height grows with the maximal velocity of the wind produced by CH1 which reflects its evolution during the period between \textit{STEREO-B} and \textit{STEREO-A} measurements. The peaks in the density distribution related to the ARCH longitudinal range are observed far from the position of the stream interface and hypothetically may be linked with streams from local sources inside the ARCH region, although it cannot be excluded that some of them can be produced also by inhomogeneity of the stream measured by instruments in different latitudinal positions.

The probable link between the solar wind and the outgoing plasma with the temperature of $\sim{1}$ MK is confirmed by the results of \opencite {Habbal10}. They analyzed the Fe charge distribution in the solar wind measured with the SWICS/\textit{ACE} in 1998 - 2009 and found a peak centered on Fe$^{8+}$, Fe$^{9+}$ and Fe$^{10+}$ ions. An iterative process based on this distribution and on the Fe ion fraction as a function of electron temperature yields a narrow peak at 1.1 MK. This value well agrees with the dominating component of the plasma temperature in outflows and coronal rays found in our case study.

\section{Summary and Conclusions}

We analyzed coronal structures seen in the 1 MK EUV iron ions lines in order to prove that coronal rays and fan loops may represent signatures of plasma outflows in the inner corona. Coronal rays known as large super-radially diverging structures above some ARs are the most probable tracers of the plasma flows in the inner corona. They may be easily distinguished from other large-scale coronal structures as super-high closed loops by their co-alignment with open magnetic field lines. Fan loops as specific small diverging structures seen at periphery of ARs are often associated with outflows and downflows.

To examine a relationship between coronal rays and fan rays with the spectroscopically detected plasma outflows, we investigated an isolated AR observed by different instruments in July -- August 2009 at the disk as well as at the limb. The EUV images of the AR were compared with magnetic field configurations calculated using the PFSS model and images of the outer corona provided by the WL coronagraphs. The main results of this study are summarized as follows.

\smallskip
1. During the period from July 25 to August 7, 2009 the AR was sequentially observed by EUVI/\textit{STEREO-B}, EIT and TESIS, EUVI/\textit{STEREO-A}.  At the beginning of this period the structure of the AR seen at the disk contained closed loops and several fan loops, but no evident rays were observed by TESIS at the eastern limb (the cause may be their poor visibility due to 3 day difference in time between the TESIS and \textit{STEREO-B} observations). On August 1 EIS detected at the disk center two compact regions of strong outflows with velocities up to 30 \kms. The western region 1 was located in the interspace between the AR and neighboring small coronal hole (CH2) with outgoing open magnetic field lines, the eastern region 2 coincided with the roots of closed magnetic loops. The EUVI/\textit{STEREO-A} image of the ARCH region near the eastern limb displayed coronal rays which were co-aligned with open magnetic field lines linking this region with the equatorial streamer. A week later, on August 7 TESIS observed coronal rays above the ARCH region at the western limb which were co-aligned with the open field lines connected with the weak streamer inclined southward. Apparently, in two last cases coronal rays were emerged from the ARCH region along open magnetic field lines. According to the EIS velocity map, it is likely that the observed coronal rays represented direct signatures of the outflow from region 1, though it is difficult to confirm it straightly due to low resolution of the magnetic field map and the EUVI images at the limb. The outflow from region 2, probably, propagated along closed loops, although it cannot be excluded that PFSS did not disclose small-sized island of open fields in this region.

2. Fan loops expanded from both outflow regions along closed magnetic field lines in the limits of EIS FOV. The EIS intensity map shows that the fan loops were the brightest in the Fe~{\sc x} line and gradually disappeared with growing ion number. Scanning of the velocity map across the outflow region 1 and adjacent fan loops has shown that in the Fe~{\sc x} line the velocity starting as negative in the outflow area then became positive in the fan loops displaying downflow.  In the lines of higher ions the velocity changed from negative in region 1 to zero in the fan loops. The total outflow flux integrated over the EIS FOV in the Fe~{\sc x} line was approximately in balance with that of downflow. In the lines of higher ions from Fe~{\sc xii} to Fe~{\sc xv} the balance was strongly shifted to outflow. The DEM analysis has shown that plasma in the outflows and in the adjacent fan loops had similar dominating temperature $\sim$1 MK which is typical also for coronal rays. These results agree with the conclusions of \inlinecite{Ugarte09} and \inlinecite{Warren11}, that fan structures are parts of cool loops indicating downflows,  whereas outflows are observed primarily in the Fe {\sc xi} to Fe {\sc xv} emission lines. The TESIS movie in the supplement clearly shows, that when the Sun rotated, the fan loops observed at the disk gradually transformed into coronal rays at the limb. Topologically, fan loops may be linked with coronal rays as the lateral part of the same magnetic structure and possibly can be regarded as secondary signatures of outflows propagated along open field lines or high closed loops.

3. We obtained some new results in the study of properties of outflows which need further investigations. The histograms of velocity distributions in Figure~\ref{fig5} show that in both regions the mean velocity grows from Fe~{\sc x} to Fe~{\sc xiii} ions and decreases from Fe~{\sc xiii} to Fe~{\sc xv} ions. This behavior differs from the progressive increase of the outflow velocity with the ion number observed earlier by \inlinecite{DelZanna08}. The basements of outflows at the photosphere were allocated in regions of moderate magnetic field of both polarities with the strength less than 200 Gs. It differs from the result of \inlinecite{Marsch04}, who found that in the case of flows along closed AR loops the outward flows corresponded to the strongest fields more than 200 G in the region of negative field polarity while the inward mass flows were associated with positive polarity. Interpretation of these results may be important for understanding of physical conditions favorable to appearance of outflows and their modeling.

4. Using the WSA map of derived coronal holes, we found that the ARCH was the source region produced the slow solar wind at the longitudes between 255 and 285$^0$ at the SS. With the help of the simple ballistic model we compared the associated solar wind data obtained by \textit{STEREO-B}, \textit{ACE}, \textit{WIND} and \textit{STEREO-A} spacecrafts at three times during the period from July 30 to August 9. At all times the solar wind was slow with a velocity between 300 and 420 \kms which well agrees with the WSA model prediction. The values of the O$^{7+}$/O$^{6+}$ and Fe/O ratios measured by \textit{ACE}  were also typical for the interface region between AR and CH. However, it was not possible to identify exactly sources of the solar wind within the ARCH with the observed outflows, because the WSA model provide locations of the source regions with some uncertainty and cannot calculate distribution of the flux density at SS. The quantitative relationship between local outflows at the Sun and the solar wind can be established by a detailed MHD modeling with consideration of quasi-stationary plasma flows. The results of our study provide sufficient starting point for such work.

\smallskip

In conclusion, alongside with development of new solar wind models, a detailed study of plasma flows and their coronal signatures using the methods of high resolution imaging EUV spectroscopy allows to reveal the primary sources of the solar wind and gives very important information about their appearance and properties.

\begin{acks}
Authors are very grateful to Dr.~Y.-M. Wang for providing and commenting the WSA map of the derived coronal holes for CR 2086, to Nariaki ~Nitta and Andrey ~Zhukov for useful discussions.
Many thanks to anonymous referee for valuable remarks and advises which helped to significantly improve the paper.
{\it Hinode} is a Japanese mission developed and launched by ISAS/JAXA,
collaborating with NAOJ as a domestic partner, NASA and STFC (UK) as
international partners. Scientific operation of the {\it Hinode} mission
is conducted by the {\it Hinode} science team organized at ISAS/JAXA. This
team mainly consists of scientists from institutes in the partner
countries. Support for the post-launch operation is provided by JAXA
and NAOJ (Japan), STFC (U.K.), NASA (U.S.A.), ESA, and NSC (Norway).
We are grateful to NASA, ESA and the {\it TRACE} and {\it STEREO} teams for their open data policy.
{\it SOHO} is a project of international cooperation between ESA and NASA.
The {\it CORONAS-Photon}/TESIS data were provided by the team of the Laboratory of X-ray Astronomy of the Sun of the P.N. Lebedev Physical Institute of Russian Academy of Sciences. We thank the {\it ACE} SWEPAM and SWICS instrument teams and the {\it ACE} Science Center for providing the {\it ACE} data. The WIND SWE experiment is a collaborative effort of Goddard Space Flight Center (GSFC), University of New Hampshire (UNH), and Massachusetts Institute of Technology (MIT).
This work was supported by the FP-7 Projects of the European Commission N 218816 SOTERIA, N 284461 eHEROES  and the {\it PROBA2} Guest Investigation Grant. SWAP is a project of the Centre Spatial de Liege and the Royal Observatory of Belgium funded by the Belgian Federal Science Policy Office (BELSPO).
The work was partially supported by the Grant 11-02-01079 of the Russian Foundation for Fundamental Research and the Grant OFN-15 of the Russian Academy of Sciences.
\end{acks}

\end{article}
\end{document}